\documentclass{aa} 

\usepackage[figuresright]{rotating}
\usepackage{graphicx}

\usepackage{txfonts}

\begin{document}
   \title{CCD BV survey of 42 open clusters}
   \author{G. Maciejewski \and  A. Niedzielski}
   \offprints{G. Maciejewski}
   \institute{Centrum Astronomii Uniwersytetu Miko{\l}aja Kopernika, ul. Gagarina\,11, Pl-87100 Toru\'n, Poland\\ \email{gm@astri.uni.torun.pl}             }
   \date{Received 10 January 2006 / Accepted 10 January 2006}
   \abstract
 {}
 {We present results of a photometric survey whose aim was to derive structural and astrophysical parameters for 42 open clusters. While our sample is definitively not representative of the total open cluster sample in the Galaxy, it does cover a wide range of cluster parameters and is uniform enough to allow for simple statistical considerations.}
 {BV wide-field CCD photometry was obtained for open clusters for which  photometric, structural, and dynamical evolution parameters were determined. The limiting and core radii were determined by  analyzing  radial density profiles. The ages, reddenings, and distances were obtained from the solar metallicity isochrone fitting. The mass function was used to study the dynamical state of the systems, mass segregation effect and to  estimate the total mass and number of cluster members.}
 {This study reports on the first determination of basic parameters for 11 out of 42 observed open clusters. The angular sizes  for the majority of the observed clusters appear to be several times larger than the catalogue data indicate. The core and limiting cluster radii are correlated and the latter parameter is 3.2 times larger on average. The limiting radius increases with the cluster's mass, and both the limiting and core radii decrease in the course of dynamical evolution. For dynamically not advanced  clusters, the mass function slope is similar to the universal IMF slope. For more evolved systems, the effect of evaporation of low-mass members is clearly visible. The initial mass segregation is present in all the observed young clusters, whereas the  dynamical mass segregation appears in clusters older than about $\log(age)=8$. Low-mass stars are deficient in the cores of clusters older than $\log(age)=8.5$ and not younger than one relaxation time.  
} 
 {}
  \maketitle

%

\section{Introduction}\label{intro}

Open clusters are not trivial stellar systems, and their dynamical evolution is not yet fully understood. Most of them are not very populous assemblages of a few hundred stars. The least massive clusters do not last longer than a few hundred Myr (Bergond et al. \cite{bergond01}). The dynamics of more massive and populous clusters is driven by internal forces to considerable degree,  which leads to evaporation of low-mass members and to a mass segregation effect. Moreover, cluster member stars incessantly evolve along stellar evolution paths, which makes an open cluster a vivid system evolving in time; hence, star clusters are considered  excellent laboratories  of stellar evolution and stellar-system dynamics (Bonatto \& Bica \cite{bonatto05}).

To obtain a complete picture of a cluster, it is necessary to study not only its most dense region (center) but also the expanded and sparse coronal region (halo).   As wide-field CCD imaging of open clusters is usually difficult, the majority of studies published so far are based on observations of the central, most populous, and relatively dense  core region. Nilakshi et al. (\cite{nilakshi02}) have presented the first, to our knowledge, results of an extensive study of spatial structure of 38 rich open clusters  based on star counts performed on images taken from the Digital Sky Survey (DSS). Bonatto \& Bica (\cite{bonatto05}) and Bica \& Bonatto (\cite{bica05})   analyzed over a dozen open clusters in detail using 2MASS photometry. In the former paper, the possible existence of a fundamental plane of several open clusters parameters was suggested. More recently, Sharma et al.~(\cite{sharma06})  published results of studies concerning cores and coronae evolution of nine open clusters based on projected radial profiles analysis.     

In this paper a sample of 42 northern open clusters of linear diameters, distances, ages, and number of potential members from a wide range is  investigated in detail based on wide-field BV CCD photometry. The basic parameters and CCD photometry  of 11 clusters were obtained for the first time.  

This paper is organized as follows. In Sect.~\ref{observations} the sample selection, observations, and data reduction are described.   In Sect.~\ref{Radial_structure} the radial structure of clusters under investigation is presented based on star counts.   Results of color-magnitude-diagram fitting are given in Sect.~\ref{CMD_sec}.   The mass functions of target clusters are analyzed in Sect.~\ref{MF_sec}.   The obtained photometric parameters for individual clusters under investigation and the reliability of the results are discussed in Sect.~\ref{results}, while in Sect.~\ref{discussion} the relations between structural and dynamical parameters are presented and discussed.   Sect.~\ref{conclusions} contains the final conclusions.  


\section{Observations and reduction}\label{observations}


%
\begin{table}
\caption{List of observed open clusters with redetermined equatorial and Galactic coordinates.}
\label{table1} 
\centering 
\begin{tabular}{l c c c} 
\hline\hline 

Name & Coordinates J2000.0 & $l$ & $b$ \\    
 & ($hhmmss$$\pm$$ddmmss$) & $(^\circ)$ & $(^\circ)$ \\ 
\hline 
King 13 & 001004+611215 & 117.9695 & $-1.2683$\\
King 1 & 002204+642250 & 119.7626 & 1.6897\\
King 14 & 003205+630903 & 120.7486 & 0.3612\\
NGC 146 & 003258+632003 & 120.8612 & 0.5367\\
Dias 1 & 004235+640405 & 121.9639 & 1.2130\\
King 16 & 004345+641108 & 122.0949 & 1.3263\\
Berkeley 4 & 004501+642305 & 122.2377 & 1.5216\\
Skiff J0058+68.4 & 005829+682808 & 123.5814 & 5.6060\\
NGC 559 & 012935+631814 & 127.2008 & 0.7487\\
NGC 884 & 022223+570733 & 135.0659 & $-3.5878$\\
Tombaugh 4 & 022910+614742 & 134.2071 & 1.0815\\
Czernik 9 & 023332+595312 & 135.4172 & $-0.4869$\\
NGC 1027 & 024243+613801 & 135.7473 & 1.5623\\
King 5 & 031445+524112 & 143.7757 & $-4.2866$\\
King 6 & 032750+562359 & 143.3584 & $-0.1389$\\
Berkeley 9 & 033237+523904 & 146.0621 & $-2.8275$\\
Berkeley 10 & 033932+662909 & 138.6158 & 8.8785\\
Tombaugh 5 & 034752+590407 & 143.9374 & 3.5924\\
NGC 1513 & 040946+492828 & 152.5955 & $-1.6243$\\
Berkeley 67 & 043749+504647 & 154.8255 & 2.4896\\
Berkeley 13 & 045552+524800 & 155.0851 & 5.9244\\
Czernik 19 & 045709+284647 & 174.0986 & $-8.8321$\\
Berkeley 15 & 050206+443043 & 162.2580 & 1.6187\\
NGC 1798 & 051138+474124 & 160.7028 & 4.8463\\
Berkeley 71 & 054055+321640 & 176.6249 & 0.8942\\
NGC 2126 & 060229+495304 & 163.2169 & 13.1294\\
NGC 2168 & 060904+241743 & 186.6426 & 2.2061\\
NGC 2192 & 061517+395019 & 173.4298 & 10.6469\\
NGC 2266 & 064319+265906 & 187.7759 & 10.3003\\
King 25 & 192432+134132 & 48.8615 & $-0.9454$\\
Czernik 40 & 194236+210914 & 57.4762 & $-1.1003$\\
Czernik 41 & 195101+251607 & 62.0054 & $-0.7010$\\
NGC 6885 & 201140+263213 & 65.5359 & $-3.9766$\\
IC 4996 & 201631+373919 & 75.3734 & 1.3158\\
Berkeley 85 & 201855+374533 & 75.7257 & 0.9812\\
Collinder 421 & 202310+414135 & 79.4299 & 2.5418\\
NGC 6939 & 203130+603922 & 95.8982 & 12.3012\\
NGC 6996 & 205631+443549 & 85.4401 & $-0.5039$\\
Berkeley 55 & 211658+514532 & 93.0267 & 1.7978\\
Berkeley 98 & 224238+522316 & 103.8561 & $-5.6477$\\
NGC 7654 & 232440+613451 & 112.7998 & 0.4279\\
NGC 7762 & 234956+680203 & 117.2100 & 5.8483\\
\hline                                   
\end{tabular}
\end{table}
%

In this survey the cluster diameter and location on the sky were the main criteria of target selection.  In the first step  we need the \textit{New catalog of optically  visible open clusters and candidates} by Dias et al. (\cite{dias02}) to select Galactic clusters  with apparent diameters ranging from 5 to 20 arcmin and a declination larger than $+10^\circ$. The former limitation guaranteed that the entire cluster with its possible extended halo would fit in the instrument's field of view. The latter one comes from the observatory location and eliminates  potential targets that cannot be observed at elevations higher than $45^\circ$. We found 295 open clusters fulfilling these criteria.

In the second step, all small and relatively populous clusters were rejected from the sample. In these clusters stellar images are   blended, making some portion of stars undetectable due to a considerable seeing of about 5\arcsec (FWHM) at the observing location.  To avoid poorly  populated objects, hardly distinguishable  from the stellar background, the minimal number of potential cluster members was set for 20. Moreover, to obtain at least 2 mag of the main sequence coverage, only clusters for which the brightest stars were brighter than 16 mag in V were selected, since the limiting magnitude was estimated as  18.5--19.5 mag. All of the selected clusters were also visually inspected on  DSS images.  

Finally, the sample of 62 open clusters was adopted. We preferred previously unstudied open clusters, for which no basic parameters were available in the literature, and these clusters were observed with higher priority. In this paper we present results for 42 open clusters, which are listed in Table~\ref{table1}. 

The collected photometric data for 20 unstudied clusters deny their cluster nature, suggesting that they constitute only an accidental aggregation of stars on the sky. These objects will be discussed in a forthcoming  paper (Maciejewski \& Niedzielski 2007, in preparation)  where extensive, detailed analysis of every object will be presented.

\subsection{Observations}

Observations were performed with the 90/180 cm Schmidt-Cassegrain Telescope located at the Astronomical Observatory of the Nicolaus Copernicus University in Piwnice near Toru\'n, Poland. A recently upgraded telescope was used in Schmidt imaging mode with a correction plate with a 60 cm diameter and a field-flattening lens mounted near the focal plane to compensate for the curvature typical of Schmidt cameras.  

The telescope was equipped  with an SBIG STL-11000 CCD camera with a KAI-11000M CCD detector (4008 $\times$ 2672 pixels $\times$ 9 $\mu$m). The field of view of the instrument was 72 arcmin in declination and 48 arcmin in right ascension with the scale of 1.08 arcsec per pixel. The camera was equipped with a filter wheel with standard UBVR Johnson-Cousins filters. The  $2 \times 2$  binning was used to increase the signal-to-noise ratio. 

Observations were carried out between September  2005 and  February 2006 (see Table~\ref{table2} for details). A set of 4 exposures in B and V filters was acquired for each program field: 2 long (600 s) and 2 short (60 s) exposures in every filter. For open clusters containing very bright stars, 2 extra very short (10 s) exposures in each filter were obtained. One of Landolt's (\cite{landolt92}) calibration fields was observed several times during each night, at wide range of airmasses. The field was observed between succeeding program exposures, in practice every hour.

\subsection{Data reduction and calibration}


%
\begin{table}
\caption{Calibration ($a_{V}$, $a_{B}$, $b_{V}$, $b_{B}$, $a_{(V-B)}$, and $b_{(V-B)}$) and atmospheric extinction ($k_{V}$ and $k_{B}$) coefficients for individual nights.}
\label{table2} 
\centering 
\begin{tabular}{l c c c c} 
\hline\hline 

Date& $k_{V}$  & $a_{V}$ & $a_{B}$ & $a_{(B-V)}$ \\    
 & $k_{B}$ & $b_{V}$ & $b_{B}$ & $b_{(B-V)}$ \\ 

\hline 
05.09.2005 & 0.3072 & -0.0922 & 0.1876 & 1.2799 \\
 & 0.4648 & 20.4407 & 20.5982 & 0.1573 \\
06.09.2005 & 0.4614 & -0.0886 & 0.1825 & 1.2694 \\
 & 0.6387 & 20.4518 & 20.5750 & 0.1230 \\
07.09.2005 & 0.5689 & -0.0782 & 0.1775 & 1.2579 \\
 & 0.6790 & 20.5841 & 20.5501 & -0.0349 \\
08.09.2005 & 0.5346 & -0.0997 & 0.1752 & 1.2749 \\
 & 0.7134 & 20.5139 & 20.6531 & 0.1394 \\
04.10.2005 & 0.1804 & -0.1030 & 0.1754 & 1.2784 \\
 & 0.3136 & 20.1073 & 20.2526 & 0.1452 \\
05.10.2005 & 0.2448 & -0.0990 & 0.1946 & 1.2935 \\
 & 0.3767 & 20.2757 & 20.3897 & 0.1140 \\
06.10.2005 & 0.3447 & -0.0911 & 0.1682 & 1.2593 \\
 & 0.4640 & 20.4129 & 20.4985 & 0.0855 \\
07.10.2005 & 0.3621 & -0.0975 & 0.2251 & 1.3227 \\
 & 0.4879 & 20.4133 & 20.3996 & -0.0137\\
08.10.2005 & 0.1810 & -0.0830 & 0.1906 & 1.2736 \\
 & 0.3014 & 20.1319 & 20.2151 & 0.0832 \\
23.02.2006 & 0.3968 & -0.1023 & 0.1458 & 1.2642 \\
 & 0.7820 & 20.9667 & 20.7216 & 0.5354 \\
26.02.2006 & 0.1911 & -0.1281 & 0.1142 & 1.2512 \\
 & 0.3658 & 21.0104 & 21.2547 & 0.2419 \\ 
\hline 
\end{tabular}
\end{table}

%

The collected observations were reduced with the software pipeline developed for the Semi-Automatic Variability Search\footnote{http://www.astri.uni.torun.pl/\~{}gm/SAVS} sky survey (Niedzielski et al. \cite{niedzielski03}, Maciejewski \& Niedzielski \cite{maciejewski05}). CCD frames were processed with a standard  procedure including debiasing, subtraction of dark frames, and flat-fielding. 

The instrumental coordinates of stars were transformed into equatorial ones based on positions of stars brighter than 16 mag extracted from the Guide Star Catalog (Lasker et al.~\cite{lasker90}). The instrumental magnitudes in B and V bands were corrected for atmospheric extinction and then transformed into the standard system.  

The preliminary analysis, including determining the width of stellar profiles, calculating of the atmospheric extinction coefficients in both filters, and determining the transformation equations between instrumental magnitudes and the standard ones, was performed based on observations of Landolt  fields. The mean FWHM of the stellar profiles was calculated for each Landolt field  frame acquired during one night. The aperture radius used for photometric measurements was calculated as $3\sigma$ of the maximum mean FWHM obtained from the Landolt field observed  during a night, and in practice it was  between 6 and 8 arcsec. 

The atmospheric extinction coefficients $k_{V}$ and $k_{B}$  were determined for each night from  6--8 observations of the adopted  Landolt field at airmasses $X$  between 1.6 and 3.2. Typically more than 1000 stars in V and 750 in B were detected in every Landolt field  frame  and the extinction coefficient in a given filter was determined for each star from changes in its raw instrumental magnitudes with $X$. The median value was taken as the one best representing a night. The values of the atmospheric extinction coefficients for individual nights are listed in Table~\ref{table2}.

The raw instrumental magnitudes $b_{raw}$, $v_{raw}$ of stars in the Landolt field were corrected for the atmospheric extinction, and instrumental magnitudes outside the atmosphere $b$, $v$ were calculated as
\begin{equation}
 b=b_{raw}-k_{b}X\, , \;
\end{equation}
\begin{equation}
     v=v_{raw}-k_{v}X\, . \;
\end{equation}
Next, the mean values of instrumental magnitudes outside the atmosphere were calculated for every star. In every Landolt  field there were about 30 standard stars that were used to determine coefficients in the calibration equations of the form:
\begin{equation}
 V-v=a_{V}(b-v)+b_{V}\, , \;
\end{equation}
\begin{equation}
     B-b=a_{B}(b-v)+b_{B}\, , \;
\end{equation}
\begin{equation}
     B-V=a_{(V-B)}(b-v)+b_{(V-B)}\, , \; 
\end{equation}
where $B$, $V$ are standard magnitudes and $b$, $v$ are the mean instrumental ones corrected for the atmospheric extinction. The detailed list of transformation coefficients for each night is presented in Table~\ref{table2}. 

The final list of stars observed in all fields contains equatorial coordinates (J2000.0), V magnitude, and (B--V) color index. The  files with data for  individual open clusters are available on the survey's web site\footnote{http://www.astri.uni.torun.pl/\~{}gm/OCS}.


\section{Radial structure}\label{Radial_structure}

Analysis of the radial density profiles (RDP) is a commonly used method for investigating cluster structure. It loses information on 2-dimensional cluster morphology  but it provides a uniform description of its
structure with a few basic parameters instead. Defining the cluster's center is essential for the
RDP analysis. Since the coordinates of clusters as given in Dias et al. (\cite{dias02}) were found in several cases to be different from the actual ones, we started with redetermination of  the centers for all the open clusters in our sample.

\subsection{Redetermination of central coordinates}


%
\begin{table}
\begin{minipage}[t]{\columnwidth}
\caption{Structural parameters obtained from the King profile fit.}
\label{table3} 
\centering 
\renewcommand{\footnoterule}{}  
\begin{tabular}{l c c c c} 
\hline\hline 

Name & $r_{lim}$ & $r_{core}$ & $f_{0}$ & $f_{bg}$  \\
    & (\arcmin)    &  (\arcmin)  &  $\left(\frac{\mathrm{stars}}{\mathrm{arcmin}^{2}}\right)$   &  $\left(\frac{\mathrm{stars}}{\mathrm{arcmin}^{2}}\right)$   \\
(1) & (2) & (3) & (4) & (5) \\

\hline                        
King 13 & 11.8 & 3.3$\pm$0.3 & 5.02$\pm$0.27 & 3.09$\pm$0.11 \\
King 1 & 12.3 & 2.1$\pm$0.1 & 5.16$\pm$1.92 & 1.76$\pm$0.05 \\
King 14 & 9.0 & 2.3$\pm$0.4 & 2.84$\pm$0.31 & 4.97$\pm$0.09 \\
NGC 146 & 2.7 & 1.2$\pm$0.3 & 5.61$\pm$0.80 & 4.87$\pm$0.16 \\
Dias 1 & 2.3 & 0.3$\pm$0.1 & 13.29$\pm$6.33 & 2.96$\pm$0.07 \\
King 16 & 8.8 & 1.9$\pm$0.2 & 4.70$\pm$0.27 & 3.25$\pm$0.06 \\
Berkeley 4 & 3.1 & 1.3$\pm$0.3 & 2.84$\pm$0.40 & 3.57$\pm$0.09 \\
Skiff J0058+68.4 & 10.9 & 3.8$\pm$0.4 & 3.66$\pm$0.20 & 2.04$\pm$0.09 \\
NGC 559 & 14.5 & 2.3$\pm$0.2 & 6.72$\pm$0.32 & 2.38$\pm$0.09 \\
NGC 884 & 10.1 & 5.8$\pm$1.3 & 1.08$\pm$0.11 & 0.94$\pm$0.09 \\
Tombaugh 4 & 5.6 & 1.1$\pm$0.1 & 12.91$\pm$0.53 & 1.62$\pm$0.07 \\
Czernik 9 & 3.3 & 0.8$\pm$0.1 & 6.36$\pm$0.59 & 1.43$\pm$0.08 \\
NGC 1027 & 10.3 & 3.3$\pm$0.5 & 1.63$\pm$0.13 & 0.81$\pm$0.05 \\
King 5 & 10.9 & 2.4$\pm$0.2 & 5.81$\pm$0.29 & 1.96$\pm$0.09 \\
King 6 & 10.9 & 3.6$\pm$0.4 & 1.55$\pm$0.09 & 0.62$\pm$0.04 \\
Berkeley 9 & 7.3 & 1.2$\pm$0.1 & 3.88$\pm$0.18 & 1.16$\pm$0.03 \\
Berkeley 10 & 8.3 & 1.3$\pm$0.1 & 6.36$\pm$0.39 & 1.28$\pm$0.07 \\
Tombaugh 5 & 11.8 & 2.2$\pm$0.4 & 3.75$\pm$0.35 & 2.41$\pm$0.10 \\
NGC 1513 & 9.2 & 3.7$\pm$0.6 & 2.47$\pm$0.20 & 1.04$\pm$0.09 \\
Berkeley 67 & 5.2 & 1.9$\pm$0.1 & 3.53$\pm$0.16 & 0.92$\pm$0.04 \\
Berkeley 13 & 6.1 & 1.4$\pm$0.1 & 5.20$\pm$0.32 & 2.13$\pm$0.06 \\
Czernik 19 & 5.5 & 1.4$\pm$0.2 & 4.12$\pm$0.30 & 1.41$\pm$0.07 \\
Berkeley 15 & 7.6 & 1.4$\pm$0.1 & 5.74$\pm$0.28 & 1.53$\pm$0.05 \\
NGC 1798 & 9.0 & 1.3$\pm$0.1 & 9.55$\pm$0.28 & 3.14$\pm$0.05 \\
Berkeley 71 & 3.3 & 1.2$\pm$0.2 & 6.04$\pm$0.58 & 1.99$\pm$0.09 \\
NGC 2126 & 10.0 & 1.9$\pm$0.3 & 1.78$\pm$0.15 & 0.93$\pm$0.04 \\
NGC 2168 & 9.8 & 4.8$\pm$0.5 & 2.27$\pm$0.16 & 1.03$\pm$0.08 \\
NGC 2192 & 4.6 & 1.4$\pm$0.2 & 2.19$\pm$0.18 & 0.57$\pm$0.04 \\
NGC 2266 & 5.9 & 1.2$\pm$0.1 & 7.69$\pm$0.50 & 2.32$\pm$0.08 \\
King 25 & 6.3 & 2.3$\pm$0.3 & 4.93$\pm$0.34 & 1.36$\pm$0.13 \\
Czernik 40 & 8.5 & 2.3$\pm$0.3 & 8.44$\pm$0.53 & 3.40$\pm$0.18 \\
Czernik 41 & 5.6 & 1.7$\pm$0.2 & 3.96$\pm$0.28 & 1.65$\pm$0.09 \\
NGC 6885 & 8.6 & 2.4$\pm$0.3 & 2.74$\pm$0.21 & 2.66$\pm$0.08 \\
IC 4996 & 2.2 & 1.2$\pm$0.4 & 3.27$\pm$0.58 & 4.61$\pm$0.14 \\
Berkeley 85 & 5.0 & 1.5$\pm$0.2 & 4.84$\pm$0.43 & 2.94$\pm$0.09 \\
Collinder 421 & 6.1 & 1.1$\pm$0.3 & 2.67$\pm$0.46 & 0.93$\pm$0.07 \\
NGC 6939 & 15.2 & 2.2$\pm$0.1 & 6.92$\pm$0.22 & 2.62$\pm$0.06 \\
NGC 6996 & 2.1 & 0.9$\pm$0.3 & 3.58$\pm$0.78 & 3.66$\pm$0.10 \\
Berkeley 55 & 6.0 & 0.7$\pm$0.1 & 7.63$\pm$0.67 & 1.45$\pm$0.04 \\
Berkeley 98 & 4.6 & 2.1$\pm$0.3 & 4.00$\pm$0.30 & 5.40$\pm$0.08 \\
NGC 7654 & 11.2 & 5.0$\pm$0.5 & 4.39$\pm$0.21 & 3.36$\pm$0.20 \\
NGC 7762 & 9.5 & 2.4$\pm$0.2 & 5.06$\pm$0.29 & 1.45$\pm$0.08 \\
\hline                                   
\end{tabular}
\end{minipage}
\end{table}

%

Our algorithm for redetermining the central coordinates started with the  approximated coordinates taken from the compilation by Dias et al. (\cite{dias02}) or from a tentative approximate position when the  catalogue data were found to be inconsistent with the cluster position as seen on DSS charts.

To determine  the center position more accurately, two perpendicular stripes  (20 arcmin long and  3--6 arcmin wide, depending on cluster size)  were cut along declination and right ascension  starting from the approximate cluster center, and stars were counted within every stripe.  The histogram of star counts was built along each stripe with a bin size of 1.0 arcmin for the cluster with a diameter larger than 10 arcmin and 0.5 arcmin for the smaller ones. The bin with the maximum value in both coordinates  was taken as the new cluster center. This procedure was repeated until the  new center position became stable, usually a few times.  The accuracy of the new coordinates was determined by the histogram's bin size and was assumed to be 1 arc min typically. The new equatorial and Galactic coordinates are listed in Table~\ref{table1}.

\subsection{Analysis of radial density profiles} \label{sekcja3.2}

\begin{figure*}
 \centering
 \includegraphics[width=17cm]{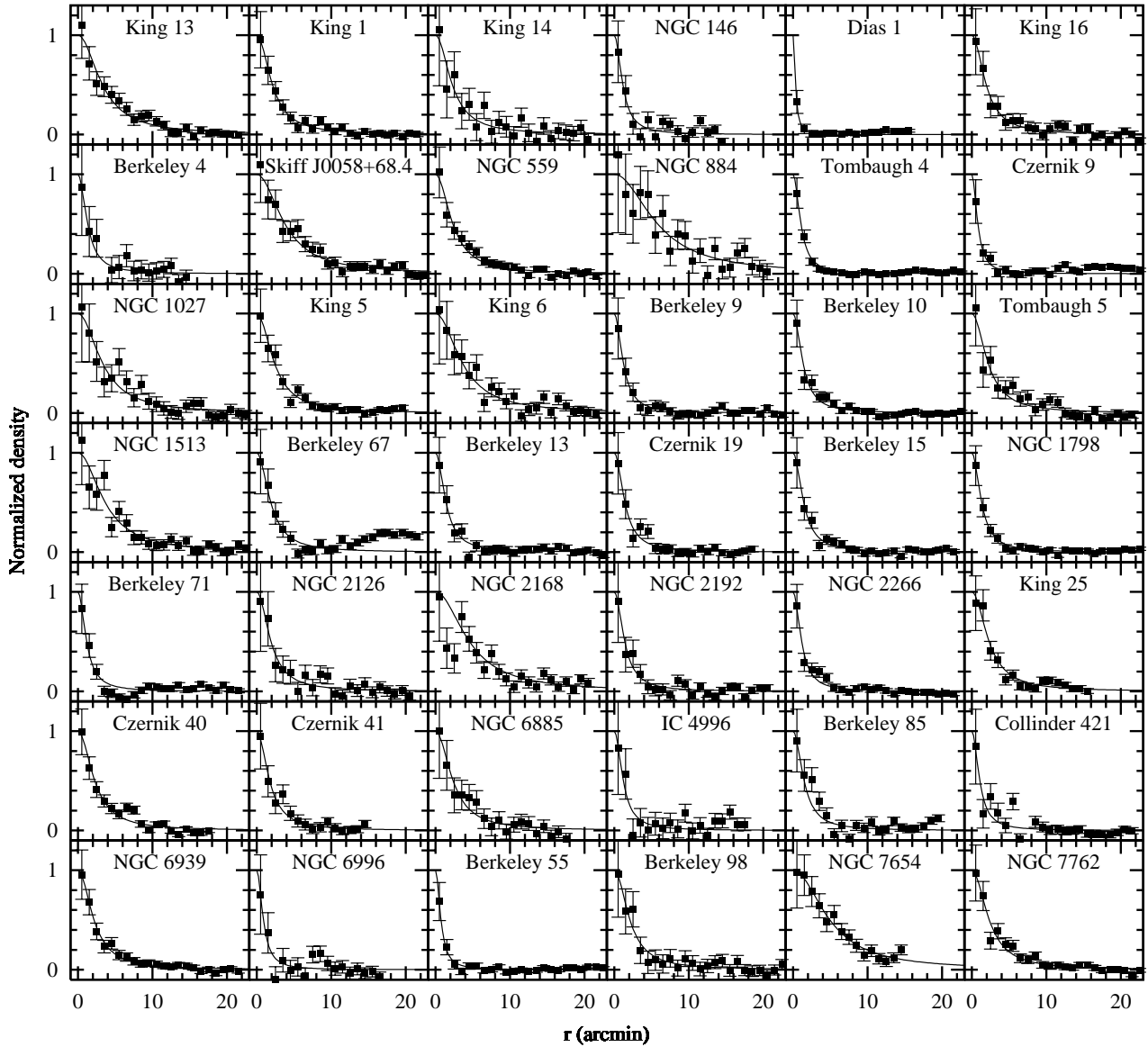}
 \caption{Radial density profiles normalized to the central value after background level subtraction. }
 \label{fig-rdp}%
\end{figure*}

The RDPs were built by calculating the mean stellar surface density in concentric rings, 1 arcmin wide, centered on the redetermined cluster center. If $N_{i}$ denotes the number of stars counted in the $i$th ring of the inner radius $R_{i}$ and outer $R_{i+1}$  the stellar surface density $\rho_{i}$ can be expressed as
\begin{equation}
 \rho_{i}=\frac{N_{i}}{\pi(R^{2}_{i+1}-R^{2}_{i})} \, . \;
\end{equation}
The density uncertainty in each ring was estimated assuming the Poisson statistics. The basic structural parameters were derived by least-square fitting the two-parameter King (\cite{king66}) surface density profile
\begin{equation}
 \rho(r)=f_{bg}+\frac{f_{0}}{1+\left(\frac{r}{r_{core}}\right)^{2}} \, ,\;
\end{equation} 
where $f_{0}$ is the central density, $f_{bg}$ the density of the stellar background in a field, and $r_{core}$ the core radius defined as the distance between the center and the point where $\rho(r)$ becomes half of the central density.  

The RDPs and the fitted King profiles are shown in Fig.~\ref{fig-rdp} where  the densities were normalized (after background density subtraction) to the central value. As one can note, all  clusters can be described by the King profile reasonably well, and even for relatively small objects, no significant systematic deviation is noticeable. The RDP of Berkeley 67 indicates the presence of a strong background gradient, so the background level was artificially straightened for the profile fitting procedure. In several cases (NGC 146, Dias 1, Berkeley 4, and NGC 7654), the RDPs were cut off at $r$ smaller than expected. That is because these clusters were located in a field centered on another open cluster and were observed serendipitously.  

The RDPs were also used to determine the limiting radius $r_{lim}$, the radius  where cluster's outskirts merge with the stellar background. This is not a trivial task and properly determining $r_{lim}$ is important for further investigations. Therefore a uniform algorithm was developed and applied to all clusters. In its first step  the boundary density level $\rho_{b}$ was calculated for every RDP as
\begin{equation}
 \rho_{b}=f_{bg}+3\sigma_{bg} \, , \;
\end{equation}
where $\sigma_{bg}$ denotes the background density error derived from the  King profile fit. Next, moving from the cluster center ($r=0$ arcmin) outwards, the first point below $\rho_{b}$ was sought. When this $i$th point was encountered, the algorithm was checked to see if farther-out points were also located below $\rho_{b}$. If this condition was fulfilled, the limiting radius was interpolated as the crossing point between the boundary density level $\rho_{b}$ and the line passing through the ($i$--1)th and $i$th points. When farther-out (at least two) points following the $i$th point were located above the boundary density level $\rho_{b}$, the algorithm skipped the $i$th point and continued seeking the next point located below $\rho_{b}$, and the procedure was repeated. As the formal error of $r_{lim}$ determination, one half of RDP bin size was taken, i.e.~0.5 arc min. Due to the limited field of view for several clusters (for instance King~13, King~16, NGC~884, NGC~1027, King~6, NGC~1513, NGC~2168, NGC~6885,  NGC~6939, NGC~7654, and NGC~7762) our determination of $r_{lim}$ may in fact represent a lower limit. The results of the RDP analysis (limiting radius $r_{lim}$, core radius $r_{core}$, central density $f_{0}$, and background level $f_{bg}$) are listed in Table~\ref{table3} in Cols. 2, 3, 4, and 5, respectively.


\section{The color--magnitude diagrams}\label{CMD_sec}


%
\begin{table*}
\caption{Astrophysical parameters obtained from isochrone fitting. }
\label{table4} 
\centering 
\begin{tabular}{l c c c c c c} 
\hline \hline 
Name & $\log(age)$ & $E(B-V)$ & $(M-m)$ & $d$ & $R_{lim}$ &  $R_{core}$ \\
     &             &  (mag)    & (mag)   & (kpc)& (pc)      & (pc)        \\
(1)  & (2)         & (3)       & (4)     & (5) & (6)       & (7)         \\
\hline 
King 13 & $8.4$ & $0.86_{-0.12}^{+0.14}$ & $15.49_{-0.58}^{+0.55}$ & $3.67_{-1.30}^{+1.37}$ & $12.6_{-4.5}^{+4.7}$ & $3.54_{-1.26}^{+1.32}$ \\
King 1 & $9.6$ & $0.76_{-0.09}^{+0.09}$ & $12.53_{-0.51}^{+0.32}$ & $1.08_{-0.33}^{+0.23}$ & $3.9_{-1.2}^{+0.8}$ & $0.67_{-0.20}^{+0.14}$ \\
King 14 & $7.0$ & $0.58_{-0.10}^{+0.10}$ & $13.75_{-0.73}^{+0.85}$ & $2.46_{-0.94}^{+1.35}$ & $6.5_{-2.5}^{+3.5}$ & $1.62_{-0.62}^{+0.89}$ \\
NGC 146 & $7.6$ & $0.56_{-0.07}^{+0.07}$ & $13.97_{-0.61}^{+0.50}$ & $2.80_{-0.89}^{+0.84}$ & $2.2_{-0.7}^{+0.7}$ & $0.94_{-0.30}^{+0.28}$ \\
Dias 1 & $7.1$ & $1.08_{-0.11}^{+0.13}$ & $14.49_{-0.57}^{+1.04}$ & $1.69_{-0.58}^{+1.21}$ & $1.2_{-0.4}^{+0.8}$ & $0.17_{-0.06}^{+0.12}$ \\
King 16 & $7.0$ & $0.89_{-0.13}^{+0.10}$ & $14.18_{-0.87}^{+0.72}$ & $1.92_{-0.85}^{+0.88}$ & $4.9_{-2.2}^{+2.3}$ & $1.09_{-0.48}^{+0.50}$ \\
Berkeley 4 & $7.1$ & $0.83_{-0.08}^{+0.08}$ & $14.53_{-0.69}^{+1.14}$ & $2.46_{-0.86}^{+1.86}$ & $2.2_{-0.8}^{+1.6}$ & $0.94_{-0.33}^{+0.71}$ \\
Skiff J0058+68.4 & $9.1$ & $0.85_{-0.13}^{+0.12}$ & $13.48_{-0.49}^{+0.57}$ & $1.48_{-0.50}^{+0.55}$ & $4.7_{-1.6}^{+1.8}$ & $1.63_{-0.55}^{+0.61}$ \\
NGC 559 & $8.8$ & $0.68_{-0.12}^{+0.11}$ & $13.79_{-0.66}^{+0.39}$ & $2.17_{-0.82}^{+0.56}$ & $9.2_{-3.5}^{+2.4}$ & $1.48_{-0.56}^{+0.38}$ \\
NGC 884 & $7.1$ & $0.56_{-0.06}^{+0.06}$ & $14.08_{-0.57}^{+0.43}$ & $2.94_{-0.87}^{+0.75}$ & $8.6_{-2.5}^{+2.2}$ & $5.00_{-1.47}^{+1.27}$ \\
Tombaugh 4 & $9.0$ & $1.01_{-0.10}^{+0.08}$ & $14.81_{-0.37}^{+0.59}$ & $2.17_{-0.58}^{+0.78}$ & $3.5_{-1.0}^{+1.3}$ & $0.67_{-0.18}^{+0.24}$ \\
Czernik 9 & $8.8$ & $1.05_{-0.14}^{+0.12}$ & $14.35_{-0.76}^{+0.36}$ & $1.66_{-0.70}^{+0.41}$ & $1.6_{-0.7}^{+0.4}$ & $0.39_{-0.17}^{+0.10}$ \\
NGC 1027 & $8.4$ & $0.41_{-0.11}^{+0.12}$ & $11.34_{-0.53}^{+0.35}$ & $1.03_{-0.34}^{+0.25}$ & $3.1_{-1.0}^{+0.7}$ & $1.00_{-0.33}^{+0.24}$ \\
King 5 & $9.1$ & $0.67_{-0.10}^{+0.09}$ & $13.82_{-0.61}^{+0.32}$ & $2.23_{-0.77}^{+0.46}$ & $7.1_{-2.4}^{+1.5}$ & $1.56_{-0.54}^{+0.32}$ \\
King 6 & $8.4$ & $0.53_{-0.11}^{+0.12}$ & $11.17_{-0.47}^{+0.55}$ & $0.80_{-0.25}^{+0.29}$ & $2.6_{-0.8}^{+0.9}$ & $0.85_{-0.27}^{+0.31}$ \\
Berkeley 9 & $9.6$ & $0.79_{-0.08}^{+0.08}$ & $12.03_{-0.52}^{+0.34}$ & $0.82_{-0.25}^{+0.18}$ & $1.7_{-0.5}^{+0.4}$ & $0.29_{-0.09}^{+0.06}$ \\
Berkeley 10 & $9.0$ & $0.71_{-0.08}^{+0.10}$ & $13.46_{-0.40}^{+0.70}$ & $1.79_{-0.46}^{+0.80}$ & $4.3_{-1.1}^{+1.9}$ & $0.70_{-0.18}^{+0.31}$ \\
Tombaugh 5 & $8.4$ & $0.80_{-0.10}^{+0.08}$ & $13.10_{-0.40}^{+0.38}$ & $1.33_{-0.37}^{+0.31}$ & $4.6_{-1.3}^{+1.1}$ & $0.85_{-0.24}^{+0.20}$ \\
NGC 1513 & $7.4$ & $0.76_{-0.18}^{+0.13}$ & $12.96_{-1.16}^{+0.76}$ & $1.32_{-0.72}^{+0.67}$ & $3.5_{-1.9}^{+1.8}$ & $1.41_{-0.77}^{+0.71}$ \\
Berkeley 67 & $9.0$ & $0.90_{-0.08}^{+0.09}$ & $13.86_{-0.37}^{+0.60}$ & $1.64_{-0.41}^{+0.61}$ & $2.5_{-0.6}^{+0.9}$ & $0.90_{-0.22}^{+0.34}$ \\
Berkeley 13 & $9.0$ & $0.66_{-0.14}^{+0.15}$ & $14.01_{-0.80}^{+1.05}$ & $2.47_{-1.07}^{+1.82}$ & $4.4_{-1.9}^{+3.2}$ & $1.02_{-0.44}^{+0.75}$ \\
Czernik 19 & $7.4$ & $0.67_{-0.08}^{+0.08}$ & $14.07_{-0.56}^{+0.73}$ & $2.50_{-0.78}^{+1.13}$ & $4.0_{-1.2}^{+1.8}$ & $1.05_{-0.32}^{+0.47}$ \\
Berkeley 15 & $8.7$ & $1.01_{-0.16}^{+0.15}$ & $15.28_{-0.46}^{+0.36}$ & $2.69_{-0.96}^{+0.71}$ & $6.0_{-2.1}^{+1.6}$ & $1.12_{-0.40}^{+0.30}$ \\
NGC 1798 & $9.2$ & $0.37_{-0.09}^{+0.10}$ & $13.90_{-0.63}^{+0.26}$ & $3.55_{-1.22}^{+0.64}$ & $9.3_{-3.2}^{+1.7}$ & $1.36_{-0.47}^{+0.25}$ \\
Berkeley 71 & $9.0$ & $0.81_{-0.08}^{+0.08}$ & $15.08_{-0.30}^{+0.65}$ & $3.26_{-0.73}^{+1.30}$ & $3.1_{-0.7}^{+1.2}$ & $1.11_{-0.25}^{+0.44}$ \\
NGC 2126 & $9.1$ & $0.27_{-0.12}^{+0.11}$ & $11.02_{-1.03}^{+0.64}$ & $1.09_{-0.52}^{+0.45}$ & $3.2_{-1.5}^{+1.3}$ & $0.61_{-0.29}^{+0.25}$ \\
NGC 2168 & $7.9$ & $0.28_{-0.16}^{+0.15}$ & $10.49_{-1.00}^{+1.14}$ & $0.84_{-0.42}^{+0.68}$ & $2.4_{-1.2}^{+1.9}$ & $1.12_{-0.56}^{+0.91}$ \\
NGC 2192 & $9.3$ & $0.04_{-0.14}^{+0.11}$ & $12.11_{-0.42}^{+0.53}$ & $2.50_{-0.81}^{+0.86}$ & $3.3_{-1.1}^{+1.1}$ & $1.02_{-0.33}^{+0.35}$ \\
NGC 2266 & $9.0$ & $0.00_{-0.09}^{+0.09}$ & $12.24_{-0.30}^{+0.50}$ & $2.81_{-0.66}^{+0.88}$ & $4.8_{-1.1}^{+1.5}$ & $0.95_{-0.22}^{+0.30}$ \\
King 25 & $8.8$ & $1.36_{-0.13}^{+0.11}$ & $15.03_{-0.93}^{+0.46}$ & $1.45_{-0.67}^{+0.44}$ & $2.7_{-1.2}^{+0.8}$ & $0.99_{-0.45}^{+0.30}$ \\
Czernik 40 & $8.9$ & $0.99_{-0.14}^{+0.13}$ & $15.52_{-0.38}^{+0.42}$ & $3.09_{-0.97}^{+0.89}$ & $7.7_{-2.4}^{+2.2}$ & $2.03_{-0.64}^{+0.59}$ \\
Czernik 41 & $8.7$ & $1.28_{-0.17}^{+0.14}$ & $14.64_{-0.87}^{+0.42}$ & $1.36_{-0.65}^{+0.40}$ & $2.2_{-1.1}^{+0.7}$ & $0.69_{-0.33}^{+0.20}$ \\
NGC 6885 & $7.1$ & $0.66_{-0.25}^{+0.14}$ & $12.06_{-1.48}^{+1.03}$ & $1.01_{-0.65}^{+0.72}$ & $2.5_{-1.6}^{+1.8}$ & $0.69_{-0.45}^{+0.50}$ \\
IC 4996 & $7.0$ & $0.58_{-0.07}^{+0.05}$ & $12.86_{-0.64}^{+0.53}$ & $1.63_{-0.53}^{+0.50}$ & $1.1_{-0.3}^{+0.3}$ & $0.57_{-0.19}^{+0.18}$ \\
Berkeley 85 & $9.0$ & $0.77_{-0.15}^{+0.14}$ & $13.61_{-0.85}^{+0.47}$ & $1.76_{-0.80}^{+0.57}$ & $2.5_{-1.2}^{+0.8}$ & $0.76_{-0.35}^{+0.25}$ \\
Collinder 421 & $8.4$ & $0.64_{-0.12}^{+0.11}$ & $12.08_{-0.48}^{+0.48}$ & $1.05_{-0.34}^{+0.33}$ & $1.8_{-0.6}^{+0.6}$ & $0.33_{-0.11}^{+0.10}$ \\
NGC 6939 & $9.1$ & $0.38_{-0.10}^{+0.18}$ & $12.15_{-0.72}^{+0.56}$ & $1.56_{-0.59}^{+0.64}$ & $6.9_{-2.6}^{+2.8}$ & $0.98_{-0.37}^{+0.40}$ \\
NGC 6996 & $8.3$ & $0.84_{-0.12}^{+0.10}$ & $13.49_{-0.66}^{+0.41}$ & $1.50_{-0.57}^{+0.40}$ & $0.9_{-0.4}^{+0.2}$ & $0.40_{-0.15}^{+0.10}$ \\
Berkeley 55 & $8.5$ & $1.74_{-0.11}^{+0.10}$ & $15.81_{-0.51}^{+0.40}$ & $1.21_{-0.39}^{+0.31}$ & $2.1_{-0.7}^{+0.5}$ & $0.26_{-0.08}^{+0.07}$ \\
Berkeley 98 & $9.4$ & $0.13_{-0.11}^{+0.11}$ & $13.26_{-0.38}^{+0.25}$ & $3.73_{-1.05}^{+0.67}$ & $5.0_{-1.4}^{+0.9}$ & $2.29_{-0.65}^{+0.41}$ \\
NGC 7654 & $7.0$ & $0.73_{-0.16}^{+0.14}$ & $13.11_{-1.12}^{+1.18}$ & $1.48_{-0.78}^{+1.24}$ & $4.8_{-2.5}^{+4.0}$ & $2.13_{-1.12}^{+1.78}$ \\
NGC 7762 & $9.3$ & $0.66_{-0.09}^{+0.08}$ & $11.52_{-0.75}^{+0.42}$ & $0.78_{-0.30}^{+0.20}$ & $2.2_{-0.8}^{+0.6}$ & $0.54_{-0.20}^{+0.14}$ \\

\hline                                   
\end{tabular}
\end{table*}

%
\begin{figure*}
 \centering
 \includegraphics[width=17cm]{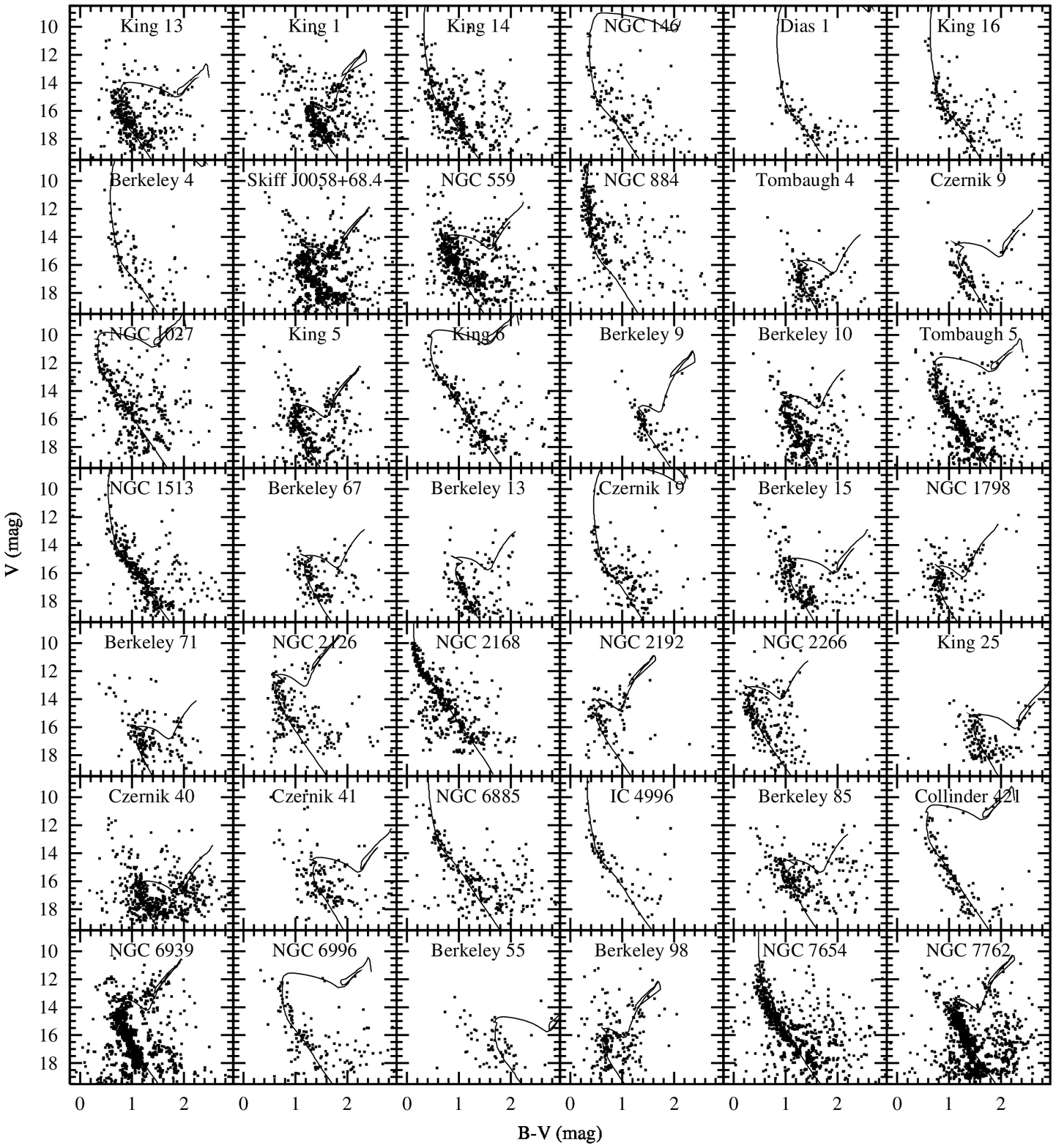}
 \caption{Decontaminated CMDs for individual clusters. The
   best-fitted isochrones are drawn with solid lines. }
 \label{fig-cmd}%
\end{figure*}


The collected $V$ and $(B-V)$ data allowed us to construct color--magnitude diagrams (CMDs) for all observed clusters. Since the field of view was wide and the majority of  clusters occurred relatively small, the CMD for the cluster region could  be decontaminated for the field stars' contribution. While in general it is impossible to point out individual cluster members based only on photometry, the contribution from field stars can be removed from the cluster CMD in a statistical manner. The algorithm applied to our data was based on ideas presented in Mighell et al. (\cite{mighell96}) and discussed in Bica \& Bonato (\cite{bica05}). 

Two separate CMDs were built: one for the cluster  and one for an offset field. The offset field was defined as a ring of the inner radius $r_{lim}$+1 arcmin and the outer radius was set as large as possible to fit within the observed CCD frame but avoiding contribution from other clusters, typically 15--19 arcmin.  Both CMDs were divided into 2-dimensional bins of $\Delta V=0.4$ mag and $\Delta(B-V)=0.1$ mag size (both values being fixed after a series of tests, as a compromise between resolution and the star numbers in individual boxes). The number of stars within each box was counted. Then the cleaned cluster CMD was built by subtracting the number of stars from the corresponding offset box from the number of stars in a cluster box. The latter number was weighted with the cluster to offset field surface ratio. Knowing the number of cluster stars occupying any given  box on clean CMD, the algorithm randomly chose the required number of stars with adequate $V$ magnitude and $B-V$ color index from the cluster field. Finally,  the list of  stars in aech cleaned cluster box was saved and used for constructing the decontaminated CMD.  

The  photometric parameters, such as distance modulus, reddening, and age of the target clusters, were derived by fitting a set of theoretical isochrones of sollar metallicity (Bertelli at al. \cite{bertelli94}) to the decontaminated CMDs. For every isochrone of a given age, a grid of $\chi^{2}$ was calculated for a number of observed distance moduli and reddenings in steps of 0.01 mag. The isochrone with the lowest $\chi^{2}$ value was chosen as the final result.

The resolution of the isochrone set was assumed to be the cluster age uncertainty, i.e. 0.1 in $\log(age)$. A map of scaled chi-square statistics $\Delta \chi^{2}$ for the best-fit isochrone was prepared to estimate the uncertainties of $E(B-V)$ and $(M-m)$. Here, $\Delta \chi^{2}$ was defined as
\begin{equation}
	\Delta \chi^{2} = \frac{\chi^{2}-\chi^{2}_{min}}{\chi^{2}_{min}/\nu} \, , \;
\end{equation}
where $\chi^{2}_{min}$ is the minimum $\chi^{2}$ and $\nu$ the number of degrees of freedom (equal 2 in this case, Burke at al. \cite{burke04}). The projection of the $\Delta \chi^{2}=1.0$ contour on the parameter axes was taken as the 1-$\sigma$ error.

The decontaminated CMDs for individual clusters are presented in Fig.~\ref{fig-cmd} where the best-fit isochrones are also shown. The parameters such as log(age), reddening, and distance modulus obtained for investigated clusters are listed in Table~\ref{table4} in Cols. 2, 3, and 4, respectively. The distances were calculated under the assumption of the total-to-selective absorption ratio of $R=3.1$ and are listed in Col. 5. The linear sizes of limiting radii $R_{lim}$ and core radii $R_{core}$ are also listed in Cols. 6 and 7, respectively.


\section{Mass functions}\label{MF_sec}

\begin{table*}
\caption{Astrophysical parameters obtained from the mass function analysis. }
\label{table5} 
\centering 
\begin{tabular}{l c c c c c c c c c} 
\hline \hline 

Name & $\chi$ & $\chi_{core}$ & $\chi_{halo}$ & $N_{evolved}$ & $M_{turnoff}$ & $N_{tot}$ & $M_{tot}$ & $N_{core}$ & $M_{core}$ \\
 & & & & (stars) & ($M_{\odot}$) & (stars) & ($M_{\odot}$) &(stars) & ($M_{\odot}$) \\
(1) & (2) & (3) & (4) & (5) & (6) & (7) & (8) & (9) & (10) \\
\hline 

King 13 & $2.23\pm1.23$ & $0.63\pm1.62$ & $2.57\pm1.02$ & $14$ & $3.36$ & $44538$ & $15598$ & $1178$ & $757$  \\
King 1 & $2.43\pm1.56$ & $-3.46\pm0.44$ & $3.58\pm1.81$ & $45$ & $1.16$ & $10520$ & $3305$ & $188$ & $177$  \\
King 14 & $1.16\pm0.38$ & $0.63\pm0.17$ & $1.35\pm0.63$ & $0$ & $14.77$ & $1436$ & $906$ & $316$ & $244$ \\
NGC 146 & $1.16\pm0.66$ & $-$ & $-$ & $0$ & $7.71$ & $1066$ & $624$ & $-$ & $-$  \\
Dias 1 & $1.07\pm0.38$ & $-$ & $-$ & $0$ & $5.27$ & $426$ & $248$ & $-$ & $-$  \\
King 16 & $1.12\pm0.37$ & $0.66\pm0.27$ & $1.26\pm0.39$ & $0$ & $11.90$ & $1084$ & $709$ & $244$ & $261$  \\
Berkeley 4 & $0.75\pm0.54$ & $-$ & $-$ & $0$ & $12.81$ & $479$ & $467$ & $-$ & $-$  \\
Skiff J0058+68.4 & $-0.38\pm0.66$ & $-0.85\pm1.54$ & $0.36\pm0.86$ & $76$ & $1.79$ & $1053$ & $851$ & $232$ & $214$  \\
NGC 559 & $1.31\pm0.43$ & $0.02\pm0.33$ & $2.09\pm0.66$ & $28$ & $1.97$ & $7286$ & $3170$ & $585$ & $395$  \\
NGC 884 & $-0.05\pm0.28$ & $-0.81\pm0.23$ & $0.49\pm0.22$ & $1$ & $15.24$ & $341$ & $1103$ & $104$ & $732$  \\
Tombaugh 4 & $0.53\pm2.51$ & $-6.03\pm2.95$ & $2.42\pm0.47$ & $6$ & $1.75$ & $3407$ & $1744$ & $102$ & $164$  \\
Czernik 9 & $1.71\pm1.58$ & $-$ & $-$ & $7$ & $2.38$ & $1424$ & $559$ & $-$ & $-$  \\
NGC 1027 & $1.51\pm0.36$ & $0.47\pm0.80$ & $1.83\pm0.39$ & $0$ & $3.34$ & $1946$ & $833$ & $127$ & $89$  \\
King 5 & $1.52\pm0.30$ & $0.06\pm0.09$ & $1.77\pm0.42$ & $22$ & $1.88$ & $5933$ & $2313$ & $417$ & $227$  \\
King 6 & $1.74\pm0.39$ & $1.44\pm0.32$ & $1.58\pm0.47$ & $0$ & $3.35$ & $1172$ & $465$ & $321$ & $141$  \\
Berkeley 9 & $1.94\pm1.27$ & $-5.58\pm1.85$ & $3.40\pm2.05$ & $4$ & $1.24$ & $2097$ & $697$ & $10$ & $14$  \\
Berkeley 10 & $1.27\pm0.55$ & $-0.66\pm0.75$ & $2.45\pm0.69$ & $12$ & $1.97$ & $2641$ & $1121$ & $67$ & $61$  \\
Tombaugh 5 & $1.31\pm0.38$ & $0.65\pm0.27$ & $1.82\pm0.22$ & $7$ & $3.33$ & $2750$ & $1287$ & $257$ & $163$  \\
NGC 1513 & $1.55\pm0.20$ & $0.54\pm0.36$ & $1.27\pm0.20$ & $0$ & $9.16$ & $2813$ & $1317$ & $365$ & $295$  \\
Berkeley 67 & $-1.66\pm1.43$ & $-2.61\pm0.86$ & $-1.02\pm3.31$ & $6$ & $1.86$ & $112$ & $140$ & $31$ & $43$  \\
Berkeley 13 & $1.87\pm0.96$ & $1.67\pm2.72$ & $2.84\pm1.29$ & $7$ & $1.92$ & $1960$ & $717$ & $983$ & $370$  \\
Czernik 19 & $1.14\pm0.18$ & $0.48\pm0.18$ & $0.87\pm0.22$ & $0$ & $9.76$ & $811$ & $502$ & $186$ & $171$ \\
Berkeley 15 & $1.10\pm1.09$ & $0.00\pm1.67$ & $1.43\pm1.30$ & $12$ & $2.21$ & $2574$ & $1230$ & $170$ & $124$ \\
NGC 1798 & $3.13\pm0.57$ & $-1.13\pm1.43$ & $4.69\pm1.29$ & $28$ & $1.74$ & $23209$ & $6932$ & $202$ & $229$ \\
Berkeley 71 & $-1.06\pm1.83$ & $-$ & $-$ & $9$ & $1.92$ & $232$ & $256$ & $-$ & $-$ \\
NGC 2126 & $1.14\pm0.41$ & $0.73\pm0.86$ & $1.20\pm0.57$ & $10$ & $1.82$ & $901$ & $395$ & $223$ & $106$  \\
NGC 2168 & $0.93\pm0.20$ & $0.82\pm0.51$ & $0.89\pm0.16$ & $0$ & $4.25$ & $1421$ & $849$ & $339$ & $216$  \\
NGC 2192 & $-4.59\pm2.12$ & $-9.50\pm3.35$ & $-3.12\pm1.14$ & $23$ & $1.54$ & $78$ & $107$ & $5$ & $8$  \\
NGC 2266 & $1.58\pm0.91$ & $0.15\pm0.52$ & $2.28\pm0.96$ & $12$ & $1.93$ & $3570$ & $1392$ & $312$ & $167$  \\
King 25 & $1.73\pm1.00$ & $0.43\pm1.38$ & $2.62\pm1.84$ & $9$ & $2.31$ & $6075$ & $2336$ & $467$ & $278$  \\
Czernik 40 & $2.41\pm0.62$ & $0.49\pm1.28$ & $3.37\pm1.13$ & $71$ & $2.13$ & $47503$ & $15790$ & $1066$ & $597$  \\
Czernik 41 & $0.77\pm0.70$ & $-2.22\pm2.83$ & $2.72\pm0.54$ & $5$ & $2.56$ & $1161$ & $635$ & $24$ & $44$  \\
NGC 6885 & $1.68\pm0.27$ & $0.70\pm0.38$ & $1.66\pm0.42$ & $0$ & $6.78$ & $711$ & $1641$ & $250$ & $141$  \\
IC 4996 & $0.87\pm0.45$ & $-$ & $-$ & $0$ & $14.47$ & $347$ & $304$ & $-$ & $-$  \\
Berkeley 85 & $1.50\pm0.73$ & $-1.86\pm0.99$ & $1.99\pm0.79$ & $10$ & $1.94$ & $4082$ & $1618$ & $40$ & $54$  \\
Collinder 421 & $1.04\pm0.37$ & $0.97\pm0.42$ & $1.32\pm0.36$ & $6$ & $3.14$ & $424$ & $233$ & $153$ & $81$  \\
NGC 6939 & $0.96\pm0.46$ & $0.19\pm0.48$ & $1.56\pm0.44$ & $48$ & $1.80$ & $5154$ & $2363$ & $786$ & $391$  \\
NGC 6996 & $1.73\pm0.57$ & $-$ & $-$ & $3$ & $3.72$ & $664$ & $277$ & $-$ & $-$  \\
Berkeley 55 & $0.91\pm1.54$ & $-1.53\pm0.81$ & $2.34\pm2.00$ & $4$ & $3.00$ & $1466$ & $795$ & $45$ & $85$  \\
Berkeley 98 & $2.52\pm1.18$ & $0.94\pm2.60$ & $3.22\pm1.35$ & $29$ & $1.41$ & $6158$ & $1967$ & $1024$ & $421$  \\
NGC 7654 & $1.42\pm0.15$ & $1.11\pm0.13$ & $1.89\pm0.37$ & $0$ & $9.23$ & $6163$ & $3133$ & $2351$ & $1491$  \\
NGC 7762 & $-0.12\pm0.33$ & $-0.35\pm0.78$ & $0.24\pm0.36$ & $28$ & $1.47$ & $1106$ & $616$ & $186$ & $113$  \\

\hline                                   
\end{tabular}
\end{table*}
%

\begin{figure*}
 \centering
 \includegraphics[width=17cm]{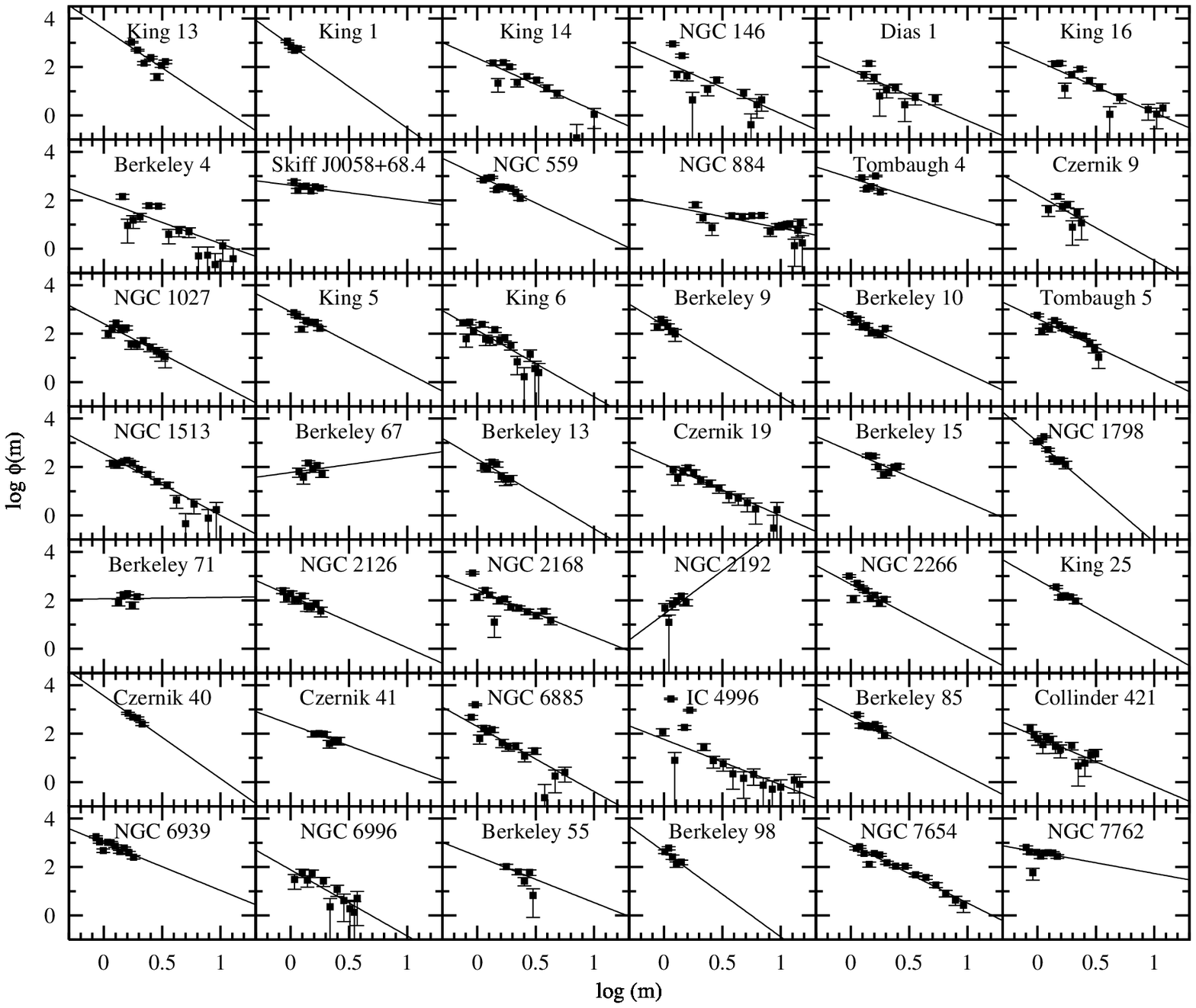}
 \caption{The mass functions for individual  clusters with the standard relation (Eq. 10) fitted (solid lines).}
 \label{fig-mf}
\end{figure*}


The first step towards deriving the cluster mass function (MF) was to build the cluster's luminosity functions (LF) for the core, halo, and overall regions separately. We used 0.5 mag bins. Another LF was built for an offset field starting at $r=r_{lim}+1$ arcmin and extending to the edge of the clean field on a frame. The LF of the offset field was subtracted, bin by bin, from every region LF, taking the area proportion into account, and this way the decontaminated LF was derived. The resulting LFs were converted into MFs using the respective isochrone. The derived mass functions $\phi(m)$ for the overall cluster region, defined as the number of stars $N$ per mass unit, are plotted as functions of stellar mass $m$ in Fig.~\ref{fig-mf}, where the standard relations of the form 
\begin{equation}
 \log\phi(m)=\log\left(\frac{dN}{dm}\right)=-(1+\chi)\log{m}+b_{0} \, , \;
\end{equation}
fitted to the data for each cluster, are also shown. The error bars were calculated assuming the Poisson statistics. The values of the MF slope parameters $\chi$ for overall clusters regions are listed in Col. 2 of Table~\ref{table5}.

This procedure was applied to objects with $r_{lim}>4'$. For smaller ones, core and halo regions were not separated to avoid small number statistics. The resulting fit  parameters $\chi_{core}$ and $\chi_{halo}$ are collected in Table~\ref{table5} in Cols. 3 and 4, respectively.  

The completeness of our photometry was estimated by adding a set of artificial stars to the data. It was defined as a ratio of the number of artificial stars recovered by our code and the number of artificial stars added. To preserve the original region crowding, the number of artificial stars was limited to 10\% of the number of actually detected stars found in the  original images within a given magnitude bin. The completeness factor was calculated for every magnitude bin. The obtained completeness factor was close to 100\% for stars brighter than 17 mag in all clusters and decreased for fainter stars more or less rapidly depending on the stellar density in a given field. The faint limit of the LM was set individually for each field after careful inspection of the observed faint-end range (typically 18--19 mag) with the completeness factor lower than 50\%. 

The derived cluster parameters  allowed us to estimate the total mass $M_{tot}$, total number of stars $N_{tot}$, core mass $M_{core}$, and  the number of stars within the core  $N_{core}$ for each cluster. These quantities were calculated by extrapolating the  MF from the turnoff down to the H-burning mass limit of 0.08 $M_{\odot}$ using the method  described in Bica \& Bonatto (\cite{bica05}). If the value of $\chi$ was similar or greater than that of the universal initial mass function (IMF), $\chi_{IMF}=1.3\pm0.3$ (Kroupa \cite{kroupa01}), the mass function was extrapolated with given $\chi$ to the mass of 0.5 $M_{\odot}$ and then with $\chi=0.3$ down to 0.08 $M_{\odot}$. For the lower actual values of $\chi$, the MFs were extrapolated with the actual value within the entire range from the turnoff mass down to 0.08 $M_{\odot}$. The contribution of the evolved stars was included in the cluster's total mass by multiplying their actual  number $N_{evolved}$ (Col. 5 in Table~\ref{table5}) by the  turnoff mass $M_{turnoff}$ (Col. 6 in Table~\ref{table5}). The total number of cluster's stars $N_{tot}$, total cluster's mass $M_{tot}$, number of stars in the core $N_{core}$, and the core mass $M_{core}$ are given in Cols. 7, 8, 9, and 10 of Table~\ref{table5}, respectively.  

To describe the dynamical state of a cluster under investigation, the relaxation time was calculated in the form
\begin{equation}
 t_{relax}=\frac{N}{8\ln N}t_{cross}\, , \;
\end{equation}
where $t_{cross}=D/\sigma_{V}$ denotes the crossing time, $N$ is the total number of stars in the investigated region of diameter $D$, and $\sigma_{V}$ is the velocity dispersion (Binney \& Tremaine \cite{binney87}) with a typical value of 3 km s$^{-1}$ (Binney \& Merrifield \cite{binney98}). The calculations were performed separately for the overall cluster and core region.  

The cluster dynamic evolution was described by the dynamical-evolution parameter $\tau$, defined as
\begin{equation}
 \tau=\frac{age}{t_{relax}}\, , \;
\end{equation}
which was calculated for the  core and the overall cluster separately. The difference between MF slopes of the core and corona $\Delta\chi=\chi_{halo}-\chi_{core}$ can be treated as the  mass segregation measure. These quantities were used for a statistical description of  the cluster's sample properties.  


\section{Reliability of results and comparison with previous studies}\label{results}

\begin{figure}
 \includegraphics[width=\columnwidth]{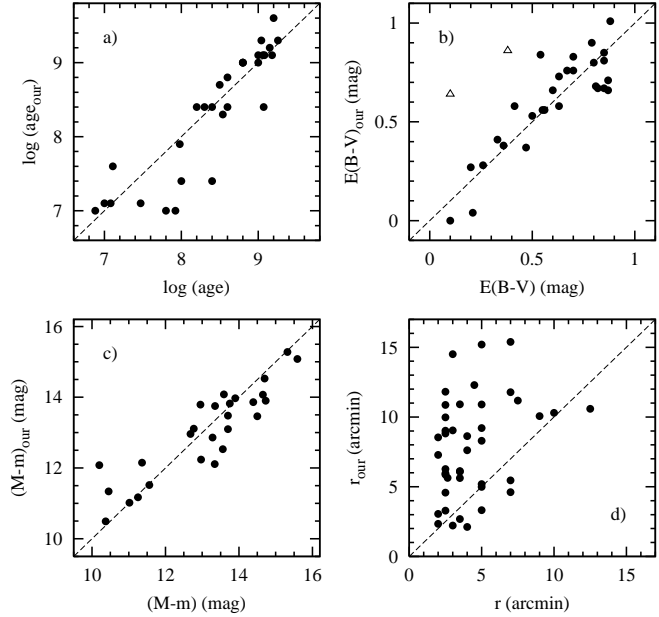}
 \caption{Reliability of the obtained results. The data obtained in this paper (vertical
      axes) are plotted versus the literature ones (horizontal axes).}
 \label{fig-comp}
\end{figure}

To test the reliability of the results of age, reddening, distance modulus, and apparent diameter determination presented in this paper, we compared them with the available catalogue data taken from the WEBDA\footnote{http:///www.univie.ac.at/webda} open cluster data base (Mermilliod \cite{mermilliod96}). Our determinations of basic cluster parameters are plotted against the
results of the previous studies of 30 clusters in Fig.~\ref{fig-comp}. The ages of clusters show excellent agreement with the literature data. The least-square-fitted linear relation for the 30 clusters is
\begin{equation}
 \log(age_{our})=(0.993\pm0.009)\log(age)
\end{equation}
with a correlation coefficient of 0.90 and fits the perfect match line within the error.

As shown in Fig.~\ref{fig-comp}b, satisfactory reliability of our $E(B-V)$ determination was also achieved. Only two clusters come significantly off  the line of perfect match. The least-square-fitted linear relation for the 28 clusters is
\begin{equation}
 E(B-V)_{our}=(0.99\pm0.04)E(B-V)
\end{equation}
with a correlation coefficient of 0.88. The distance moduli determined in this study,  Fig.~\ref{fig-comp}c, are very similar to the literature data as well. The best-fit linear relation for the 27 clusters is
\begin{equation}
 (M-m)_{our}=(0.99\pm0.01)(M-m)
\end{equation}
with the correlation coefficient of 0.90. All three relations prove that our results are reliable.

As  displayed in  Fig.~\ref{fig-comp}d, the literature values of apparent radii of open clusters are considerably (4--5 times) lower for the majority of the clusters under investigation.


\section{Statistical considerations}\label{discussion}

The sample of 42 open clusters studied in detail  within this survey is by no means complete.  Since it was defined by celestial coordinates, estimated sizes, and richness of potential objects, as well as non-availability of previous CCD studies, it is definitively not representative of the total open cluster sample in the Galaxy. However  the sample covers quite a wide range of clusters parameters and is  uniform enough to perform  simple statistical analysis.

Even though 20 clusters out of 62 covered by this survey were found not to be real does not necessarily  mean that $\sim$30\% of clusters in Dias et al. (\cite{dias02}) are doubtful. Such a high frequency of accidental star density fluctuations is definitely caused by our selection criteria.

\subsection{Limiting and core radii}

\begin{figure}
 \includegraphics[width=\columnwidth]{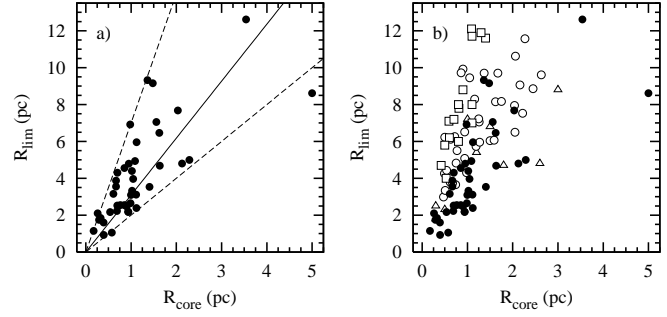}
 \caption{Relation between limiting and core radii. See text for description.}
 \label{fig-radii}
\end{figure}

From their analysis based on DSS images of 38 open clusters Nilakshi et al. (\cite{nilakshi02}) concluded that the angular size of the coronal region is about 5 times the core radius, hence $R_{lim} \approx 6 R_{core}$. Bonatto \& Bica (\cite{bonatto05})  reported a similar relation between the  core and limiting radii based on their study of 11 open clusters. Bica \& Bonatto (\cite{bica05}) used data for 16 clusters to find that  $R_{lim}=(1.05\pm0.45)+(7.73\pm0.66) R_{core}$. More recently, Sharma et al.~(\cite{sharma06}) determined core and limiting radii of 9 open clusters using  optical data and presented the relation  $R_{lim}=(3.1\pm0.5) R_{core}$ with the correlation coefficient of 0.72. The best fit obtained using the data  reported in this study gives $ R_{lim} = (3.1\pm0.2) R_{core}$ with the correlation coefficient of 0.74 (Fig.~\ref{fig-radii}~a).  Although the correlation is quite strong, $R_{lim}$ may vary for individual clusters between about $2 R_{core}$ and $7 R_{core}$ (Fig.~\ref{fig-radii}a). 

The obtained relation is quite different from the one obtained in the papers mentioned above, except for Sharma et al.~(\cite{sharma06}) who used observations gathered with a wide-field Schmidt telescope similar to ours. The field of view in surveys by Nilakshi et al. (\cite{nilakshi02}), Bonatto \& Bica (\cite{bonatto05}), and Bica \& Bonatto (\cite{bica05}) was wider with a radius of 1--2\degr. That suggests that our determinations of the limiting radius for some extensive clusters are underestimated due to a limited field of view (see Sect.~\ref{sekcja3.2}). However, it has to be pointed out that the methods of determining the cluster limiting radius differ considerably, and sometimes the adopted definition is not clear. We also note that our open clusters' size determinations differ from many in the literature at the level of angular diameters (Fig.~\ref{fig-comp}d). One of the reasons for such an inconsistency may  be the difference in the content of the cluster samples used by different authors who  frequently use non uniform photometric data. Finally, as Sharma et al.~(\cite{sharma06}) notes, open clusters appear to be larger in the near-infrared than in the optical data.

To illustrate the difference between results, we again plot our data in Fig.~\ref{fig-radii}b as in Fig.~\ref{fig-radii}a, together with the literature determinations taken from the following papers: open squares denote results from Bica \& Bonatto (\cite{bica05}) and Bonatto \& Bica (\cite{bonatto05}), open circles those of Nilakshi et al. (\cite{nilakshi02}), and open triangles Sharma et al.~(\cite{sharma06}). It is clear that data coming from  optical investigations fit each other. We also note that the literature data contain no determination of sizes for clusters smaller than 2 pc.  

\subsection{Structural parameters}

\begin{figure}
 \includegraphics[width=\columnwidth]{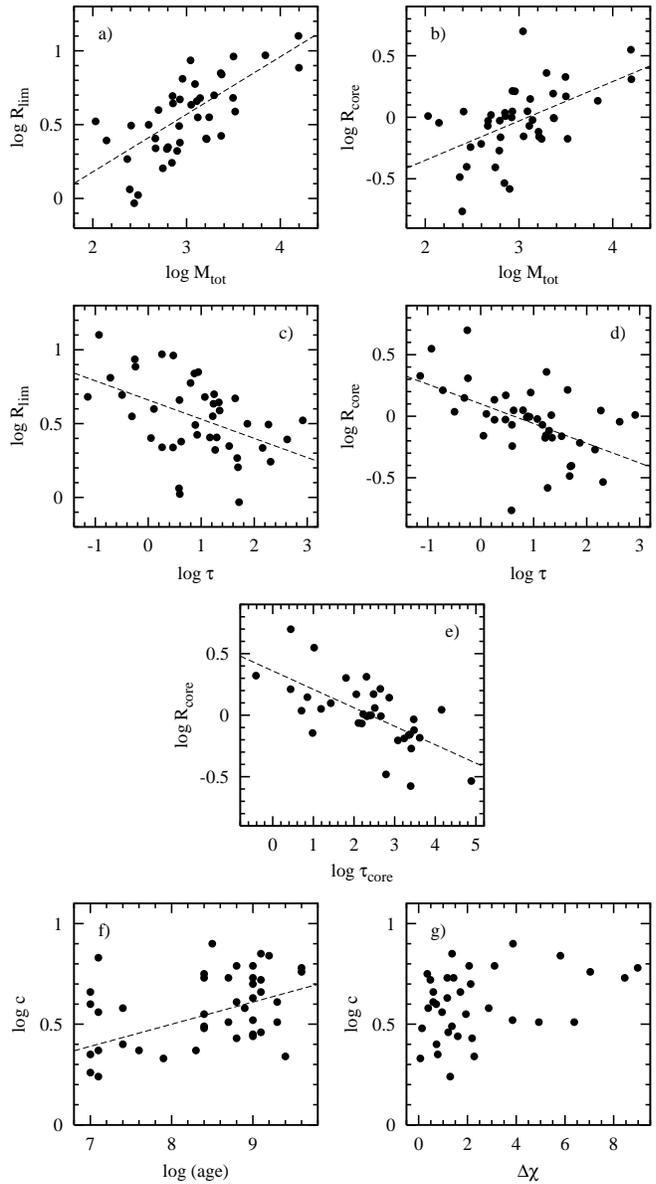}
 \caption{Relations between  structural parameters with each other. See text for details.}
 \label{fig-structural2}
\end{figure}

The data gathered within this survey show that the limiting radius correlates with the cluster's total mass (Fig.~\ref{fig-structural2}a). We obtained the relation $\log R_{lim} = (0.39\pm0.07) \log M_{tot} - (0.6\pm0.2)$ with a moderate correlation coefficient of 0.70. This result indicates that clusters with large diameters and small total masses do not form bound systems. On the other hand, small massive clusters are dissolved by the internal dynamics (Bonatto \& Bica \cite{bonatto05}). As one could expect, the core radius is also related to the cluster's total mass (Fig.~\ref{fig-structural2}b). The obtained least-square  linear relation is $\log R_{core} = (0.32\pm0.08) \log M_{tot} - (0.99\pm0.25)$ with a weak correlation coefficient of 0.53.

As shown in Figs.~\ref{fig-structural2}c and d, both radii tend to decrease in the course of the dynamical evolution. For the limiting radius, the obtained relation is $\log R_{lim} = (-0.13\pm0.04) \log \tau + (0.66\pm0.05)$ with a weak correlation coefficient of 0.47. This suggests that dynamical evolution makes a cluster smaller due to dissolving coronae. The dynamical evolution of the core radius is more visible. The least-square fitted linear relation is $\log R_{core} = (-0.16\pm0.04) \log \tau + (0.10\pm0.05)$ with a correlation coefficient of 0.51. Moreover, the core radius is moderately correlated with the dynamical-evolution parameter of the core $\tau_{core}$. The obtained relation, plotted in Fig.~\ref{fig-structural2}e, is $\log R_{core} = (-0.15\pm0.03) \log \tau_{core} + (0.36\pm0.08)$ with a correlation coefficient of 0.66. The last two relations indicate that the dynamical evolution of both, overall cluster and core, tends to reduce the core radius. No relation of $R_{lim}$ and $R_{core}$ with cluster age or mass segregation was noted in the investigated sample. 

To investigate the relative size of halos, the concentration parameter $c$, defined as $c= (R_{lim}/R_{core})$, was plotted against other parameters. The concentration parameter seems to be related to cluster age, as shown in Fig.~\ref{fig-structural2}f. For clusters younger than about $\log (age) = 9$, it tends to increase with cluster age ($\log c = (0.11\pm0.03) \log (age) - (0.38\pm0.23)$ with a correlation coefficient of 0.56). Nilakshi et al. (\cite{nilakshi02})  notes a decrease in the size of halos for older systems.

In Fig.~\ref{fig-structural2}g the concentration parameter is plotted against mass segregation parameter $\Delta \chi$. As no relation is seen, one can conclude that there is no low concentrated clusters with $\log c<0.5$ with a high value of $\Delta \chi>3$.


\subsection{Mass function slopes}

\begin{figure}
 \includegraphics[width=\columnwidth]{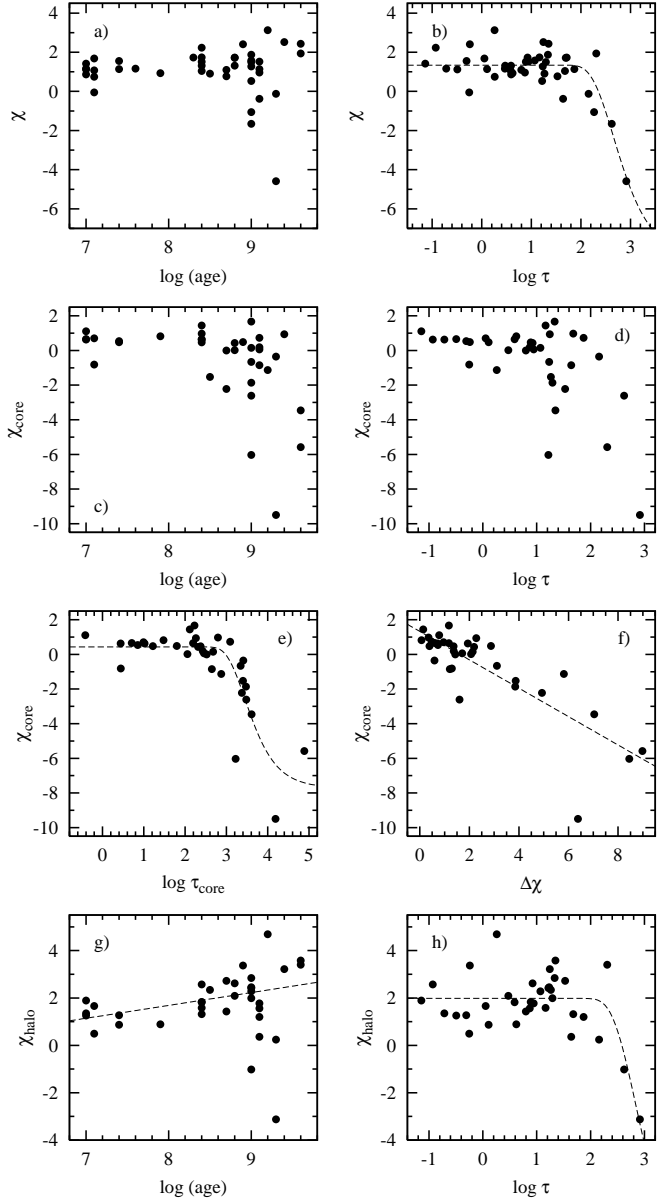}
 \caption{Relations between mass function slopes and other clusters parameters. See text for details.}
 \label{fig-dynamical}
\end{figure}

The mass function slopes of the overall cluster, core, and halo were sought for  relations with other parameters. As displayed in Fig.~\ref{fig-dynamical}a, the mass function slope of the overall cluster region and the cluster age are not strictly related. However, a deficit in low-mass members occurs in clusters older than $\log (age)=9.0$. We also investigated relations between the mass function slopes and the dynamical-evolution parameter $\tau$ -- Fig.~\ref{fig-dynamical}b. Bonatto \& Bica (\cite{bonatto05}) report a relation between the overall $\chi$ and  $\tau$ in the form of  $\chi(\tau)=\chi_{0}-\chi_{1}\exp(-\tau_{0}/\tau)$,  suggesting that the  MF slopes decrease exponentially with $\tau$. Our results confirm this relation, and the least-square fit was obtained with $\chi_{0}=1.34\pm0.13$, $\chi_{1}=9.9\pm2.3$, and $\tau_{0}=450\pm130$ (a correlation coefficient was 0.81). It is worth noting that the obtained value of $\chi_{0}$ is almost identical to $\chi_{IMF}$.

In Fig.~\ref{fig-dynamical}c the relation between $\chi_{core}$ and the $\log (age)$ is presented. It is clear that $\chi_{core}$ decreases rapidly with cluster age for clusters older than $\log (age) =8.5$. This suggests that evaporation of the low-mass members from cluster cores does not occur in clusters younger than $\log (age) =8.5$. The cores of clusters older than $\log (age) =8.5$ are dynamically evolved and deprived of low-mass stars. 

As one can see in Fig.~\ref{fig-dynamical}d, $\chi_{core}$ does not correlate with $\tau$. However, $\chi_{core}$ tends to decrease with $\tau$, which indicates that low-mass-star depleted cores appear in dynamically evolved clusters with $\log \tau > 1$.

In Fig.~\ref{fig-dynamical}e we show that $\chi_{core}$ and  $\tau_{core}$ are related, and the relation is similar to the one for $\chi$ and  $\tau$ -- $\chi_{core}=(0.44\pm0.27)-(8.1\pm1.1)\exp(-\frac{2790\pm630}{\tau_{core}})$. Such an evolution of  $\chi_{core}$ was also reported by Bica \& Bonatto (\cite{bica05}). Our fit indicates, however, that the initial $\chi_{core}(0)=0.44\pm0.27$ is much lower than the value of $1.17\pm0.23$ obtained by these authors. This suggests that $\chi_{core}$ is significantly lower than $\chi_{IMF}$ for dynamically young systems.

As displayed in Fig.~\ref{fig-dynamical}f, $\chi_{core}$ is correlated with the mass segregation parameter $\Delta \chi$. The least-square-fitted relation $\chi_{core} = (-0.83\pm0.10) \Delta \chi + (1.33\pm0.33)$ with the correlation coefficient of $-0.83$ indicates that -- as one could expect -- $\chi_{core}$ decreases with the increase in mass segregation.

Finally, in Figs.~\ref{fig-dynamical}g and h relations between $\chi_{halo}$ and  $\log (age)$ or $\tau$ were plotted, respectively. 
The MF slopes of the coronal regions of clusters younger than $\log (age) = 8.9$ tend to increase with age. That relation was marked with the least-square-fitted dashed line for which the correlation coefficient is 0.61. Bifurcation occurs for older clusters and  $\chi_{halo}$  becomes either very high or low as compared to the mean value. This suggests that the clusters were observed in different stages of dynamical evolution. In clusters with higher values of $\chi_{halo}$, the mechanism of dynamical mass segregation is more efficient than the evaporation of low-mass members from halos. Clusters with low values of $\chi_{halo}$ are dynamical evolved systems devoid of low-mass stars in the overall volume. Then, $\chi_{halo}$ seems to decrease with $\tau$ for $\log \tau > 2$. That suggests that in general the dynamical evolution of a cluster halo is driven by the dynamical evolution of the overall system  (we obtained $\chi_{halo}=(2.0\pm0.2)-(12.3\pm6.5)\exp(-\frac{700\pm350}{\tau})$ with the correlation coefficient of 0.69). It is also worth noting that the average $\chi_{halo}$ for dynamically not evolved clusters is larger than $\chi_{IMF}$.  


\subsection{Mass segregation}

\begin{figure}
 \includegraphics[width=\columnwidth]{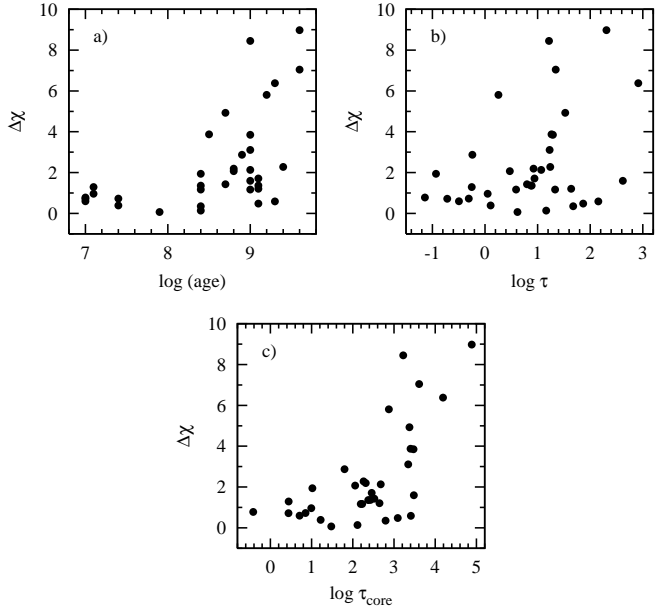}
 \caption{Evolution of the  mass segregation measure in time. See text for details.}
 \label{fig-segregation}
\end{figure}

The mass segregation $\Delta\chi$ is most prominent for clusters older than about $\log (age) =8$ (Fig.~\ref{fig-segregation}a). The mean values of  $\chi_{halo}$ and $\chi_{core}$  differ significantly for  clusters with $\log (age) <8$, and $\Delta\chi\neq0$ even for very young clusters. This suggests the existence of the initial mass segregation within the protostellar gas cloud.

We plotted $\Delta \chi$ in Fig.~\ref{fig-segregation}b as a function of the dynamical-evolution parameter $\tau$. No relation can be seen. However, one can note that strong mass segregation occurs in cluster older then their relaxation time, i.e. $\log \tau>0$. As displayed in Fig.~\ref{fig-segregation}c, mass segregation seems to be related to the core dynamical-evolution parameter $\tau_{core}$. Although a strict relation is not present, one can see that clusters with dynamically evolved cores ($\log \tau_{core} > 3$) reveal a strong mass segregation effect.


\section{Conclusions}\label{conclusions}

Wide-field CCD photometry in B and V filters was collected for 42 open clusters and the basic structural and astrophysical parameters were obtained. Eleven cluster under investigation were studied for the first time. 
 
A simple statistical analysis of our sample of open clusters leads to the following conclusions:

\begin{itemize}

\item The angular sizes of most of the observed open clusters  appeared to be several times larger than the catalogue data indicate.

\item A correlation exists between core and limiting radii of open clusters. The latter seem to be 2-7 times larger, with average ratio of 3.2. The limiting radius tends to increase with the cluster's mass. Both limiting and core radii decrease in the course of the dynamical evolution. Moreover, core radius decreases  with the core dynamical-evolution parameter.

\item The relative size of a cluster halo (in units of the core radius) tends to increase with cluster age for systems younger than $\log (age) = 9$. Among clusters with a strong mass-segregation effect, there are no systems with small halos.

\item The MF slope of the overall cluster region is related to the dynamical-evolution parameter with the  relation found in  Bica \& Bonatto (\cite{bica05}). For clusters with $\log \tau<2$, the MF slope is similar to the slope of the universal IMF. For clusters with log $\tau>2$ (older than about $\log(age)=9$), the results of evaporation of the low-mass members are seen, and $\chi$ reaches an extremely low value for clusters with log $\tau=3$.  
	
 \item The MF slope of the core region is smaller than the universal
	value even for very young clusters, while the mass
	function slope of the corona is larger.  This indicates the
	existence of the initial mass segregation. The dynamical mass
	segregation appears in clusters older than about $\log (age)=8$.

 \item A strong deficiency of low-mass stars appears in cores  of clusters
	older than $\log(age)=8.5$  and not younger than one relaxation
	time.

\end{itemize}


\begin{acknowledgements}

We thank the anonymous referee for remarks that significantly improved the paper. 
This research is partially supported by UMK grant 369-A and has made use of
the WEBDA data base operated at the Institute for Astronomy of the
University of Vienna, SIMBAD data base, as well as The Guide Star
Catalogue-II, which is a joint project of the Space Telescope Science
Institute and the Osservatorio Astronomico di Torino. Space Telescope
Science Institute is operated by the Association of Universities for
Research in Astronomy, for the National Aeronautics and Space Administration
under contract NAS5-26555. The participation of the Osservatorio Astronomico
di Torino is supported by the Italian Council for Research in Astronomy.
Additional support is provided by the European Southern Observatory, Space
Telescope European Coordinating Facility, the International GEMINI project, and the European Space Agency Astrophysics Division.

\end{acknowledgements}


\end{document}